# High throughput investigation of an emergent and naturally abundant 2D material: Clinochlore

Raphaela de Oliveira[a], Luis A. G. Guallichico[b], Eduardo Policarpo[a], Alisson R. Cadore[c], Raul O. Freitas[d], Francisco M. C. da Silva[d,e],Verônica de C. Teixeira[d], Roberto M. Paniago[a], Helio Chacham[a], Matheus J. S. Matos[b], Angelo Malachias[a], Klaus Krambrock[a], Ingrid D. Barcelos*[d]

[a] Physics Department, Federal University of Minas Gerais (UFMG), Belo Horizonte 31270-901, Minas Gerais, Brazil.
[b] Physics Department, Federal University of Ouro Preto (UFOP), Ouro Preto 35400-000, Minas Gerais, Brazil.
[c] School of Engineering, Mackenzie Presbiterian University, São Paulo 01302-907, São Paulo, Brazil.
[d] Brazilian Synchrotron Light Laboratory (LNLS), Brazilian Center for Research in Energy and Materials (CNPEM), Campinas 13083-100, São Paulo, Brazil.
[e] "Gleb Wataghin" Institute of Physics, University of Campinas (UNICAMP), Campinas 13083-859 , São Paulo, Brazil.

*ingrid.barcelos@lnls.br;

## Abstract

Phyllosilicate minerals, which form a class of naturally occurring layered materials (LMs), have been recently considered as a low-cost source of two-dimensional (2D) materials. Clinochlore $[Mg_5Al(AlSi_3)O_{10}(OH)_8]$ is one of the most abundant phyllosilicate minerals in nature, exhibiting the capability to be mechanically exfoliated down to a few layers. An important characteristic clinochlore is the natural occurrence of defects and impurities which can strongly affect their optoelectronic properties, possibly in technologically interesting ways. In the present work, we carry out a thorough investigation of the clinochlore structure on both bulk and 2D exfoliated forms, discussing its optical features and the influence of the insertion of impurities on its macroscopic properties. Several experimental techniques are employed, followed by theoretical first-principles calculations considering several types of naturally-ocurring transition metal impurities in the mineral lattice and their effect on electronic and optical properties. We demonstrate the existence of requirements concerning surface quality and insulating properties of clinochlore that are mandatory for its suitable application in nanoelectronic devices. The results presented in this work provide important informations for clinochlore potential applications and establish a basis for further works that intend to optimize its properties to relevant 2D technological applications through defect engineering.

# 1. Introduction

The research on layered materials (LMs) was initially motivated by the exfoliation of naturally occurring graphite into isolate single graphene layers [1], on which extensive studies were developed [2]. Since then, the interest in understanding the physicochemical properties of LMs and the number of investigated systems have grown. So far, more than 1800 systems with relevant two-dimensional (2D) characteristics have been computationally identified [3]. The reduction in the dimensions of a system favors the emergence of quantum effects that can modify its bulk properties in surprising ways that might contribute to technological applications [4]. Most of the attempts to find alternative LMs that are also capable of being reduced to single (1L) or few layers (FL) have been through synthesizing materials such as hexagonal boron nitride (hBN) [5] and transition metal dichalcogenides (TMDs) [6,7]. Synthetic materials can be produced in controlled environments, but some processes required for specific material classes are complex and consequently expensive. In an effort to increase the list of naturally occurring LMs that are abundant in nature and could potentially become low-cost sources of 2D materials, recent research has been carried out in the promising group of phyllosilicates minerals, which are wide bandgap insulators that can be mechanically exfoliated down to 1L and FL-flakes [8–16], and and can be embedded in van der Waals heterostructures with other LMs [17–24]. The challenge in using naturally occurring van der Waals materials for 2D applications is related to the existence of impurities and defects in their crystal lattice [25]. They can be point defects (vacancies, substitutional or interstitial impurities) or extended defects (grain boundaries, twin planes and stacking faults), strongly affecting the optoelectronic responses of the nanomaterial. Nevertheless, a deeper knowledge of such impurities and defects is essential to identify their eventual influence on a given property allowing their manipulation and tailoring to specific 2D technological applications, such as sensors [18,26,27], electronic devices [9,12,17,22–24,28] and catalysts [29,30].

Phyllosilicates are a subgroup of silicates that include clays, micas, and chlorites [31]. Most of the investigations of their 2D form have focused on muscovite mica or talc [8-23,26,27], and barely explored other minerals of this family, such as chlorites [10]. Within the chlorite group, clinochlore is one of the most abundant mineral. It is a natural layered structured crystal with the chemical formula $Mg_5Al(AlSi_3)O_{10}(OH)_8$. Previous works on its 3D-bulk form have been carried out to elucidate its optical, mechanical and electric properties [32–34]. Clinochlore has also been applied to diverse fields, such as prebiotic synthesis and polymerization of biomolecules [35], catalyst in the chemical recycling of waste plastics [36], for room-temperature synthesis of triazole compounds in water [37], and decontamination of water resources [38]. Furthermore, the geological importance of clinochlore in the study of mineral formation [39,40] and hydration/weathering processes at nanoscale [41] remains a topic of current relevance, as well as its mechanical behavior under pressure [42,43]. Our study reveals how its 2D structure and its optical properties may render this material as an alternative and naturally abundant insulator suitable for the development of 2D electronic devices, aspects poorly explored so far.

Being a natural mineral, variation may occur in both the chemical composition and crystalline structure of clinochlore specimens from different geological environments. For this reason, a systematic characterization of a clinochlore sample is carried out by several experimental techniques, including X-ray diffraction (XRD), X-ray fluorescence (XRF) and absorption near edge structure (XANES), ultraviolet reflectivity (UR), chemical analysis by energy-dispersive (EDS) and wavelength-dispersive spectroscopy (WDS), ultraviolet/visible/near infrared (UV/Vis/NIR) and Fourier-transform infrared spectroscopy (FTIR), electron paramagnetic resonance (EPR), Mössbauer and Raman spectroscopy. These techniques are used to investigate its 3D-bulk properties. Atomic force microscopy (AFM) and synchrotron infrared nanospectroscopy (SINS) are then used to analyze its few-layers characteristics. Additionally, a systematic theoretical study of the clinochlore structure is performed by density functional theory (DFT), adding the most common impurities to the probable sites of its pristine structure. Such investigation allow us to understand how the occurrence of point defects impact the macroscopic mineral properties. Our theoretical and experimental results are in good agreement and establish the basis for further studies to optimize clinochlore properties to desired technological applications in both bulk and FL forms through defect engineering. This procedure may pave the way to widening the doors for the exploration of natural LMs beyond graphene and TMDs.

## 2. Methods

### 2.1 Sample characterization

#### 2.1.1 Structural and morphological characterization

XRD measurements were carried out at an Empyrean (Malvern-Panalytical) powder diffractometer with a Cu source. The powder prepared from a fragment of the clinochlore sample was positioned in Bragg-Brentano geometry. The measurements were analyzed using the MAUD package. For elementary qualitative analysis by EDS, a Hitachi TM4000 Plus scanning electron microscope were used and previously calibrated with a Cu standard sample. The superficial layers of the bulk clinochlore sample were mechanically exfoliated for the EDS measurements. For the WDS measurements, a clinochlore sample was prepared on a resin support, polished and metallized with carbon. The quantitative microanalysis by WDS was performed for several points at different regions of the sample for a statistical quantification using a previously calibrated Jeol 8900 electron microprobe. Images by AFM were acquired using a commercial s-SNOM instrument (Neaspec GmbH) simultaneously with spectrally integrated optical near-field IR images. The tip−sample scattered light is modulated by the tip mechanical frequency ($\Omega \approx 300$ kHz, in our case) since our AFM instrument operates in semicontact (tapping) mode.

### 2.1.2 Optoeletronic and optovibracional characterization

UR measurements were performed at a toroidal grating monochromator (TGM) synchrotron beamline [44] in the energy range from 4.5 to 11.0 eV. The system pressure was maintained below $1.3 \times 10^{-7}$ mbar at room temperature. To obtain high purity photons without higher orders harmonics, two filters at the excitation were used: an ActIon quartz filter (cut-off at 7.7 eV and 200 µm thickness) to acquire the lower energy spectrum range and a Laser Optex $MgF_2$ filter (cut-off 10.9 eV and 170 µm thickness) to acquire the higher energy spectrum range. The superficial layers of the bulk clinochlore sample were mechanically exfoliated and the sample was positioned at an angle of ~15º in relation to the incident beam direction. The signal was recorded through a sodium salicylate film deposited on a glass substrate coupled to a R3809U-50 Hamamatsu microchannel plate and read in current mode with an electrometer Keithley 6514. For the Raman measurements, a confocal Raman microscope WITec Alpha 300R was used in the backscattering configuration at unpolarized under 488 nm excitation energy with a 100x objective which has a numerical aperture (NA) of 0.9. The laser power was kept constant at 10 mW during all measurements and the laser was previously aligned on a prime silicon wafer to use the Si peak at 521 $cm^{-1}$ as reference. It is worth mentioning that the Raman signal of FL-clinochlore is extremely weak and cannot be properly detected when the sample is in direct contact with the substrate. This is a common issue in phyllosilicates. For this reason, the bulk clinochlore sample was mechanically exfoliated onto $SiO_2$/Si substrates and the bulk flakes (>200 nm) chosen for the Raman measurements were flakes that were partially suspended at the edges of the substrate. The micro-FTIR measurements were carried out in a Cary 620 FTIR microscope from Agilent Technologiespowered by a blackbody thermal source. Spectral resolution was set as 8 $cm^{-1}$. Due to the low IR-transparency of the sample, single-point spectra in reflectance mode of a crystal fragment from the bulk clinochlore sample placed atop flat gold surface were recorded by a MCT detector and a 25x objective lens with approximately 400 x 400 $\mu m^2$ of spot size. The presented spectrum is an accumulation of 64 scans normalized by the spectrum of a clean Au surface as background reference. SINS was performed at the Advanced Light Source (ALS) [45] and the Brazilian Synchrotron Light Laboratory (LNLS) [46] using a commercial s-SNOM instrument (Neaspec GmbH). Briefly, IR light was coupled into an asymmetric Michelson interferometer, in which half the light passes through a beamsplitter and was focused onto a sharp metallic AFM tip. The back-scattered light from the tip-sample interaction was recombined on the beamsplitter with a reference beam from a translating mirror and focused onto a cryogenically-cooled detector. The weak near-field interaction was differentiated from the strong diffraction-limited far-field background by demodulating at harmonics of the tip-tapping frequency (~250 kHz) with a lock-in amplifier. The interferometric resulting signal was Fourier-transformed, giving the complete nano-resolved ultra-broadband near-field response, $S_n(\omega) = |S_n|\exp(i\varphi n)$, where $|S_n|$ and $\varphi_n$ were the SINS amplitude and phase, respectively, demodulated at the nth tip harmonic. All SINS spectra here were given by n = 2, i.e, $S_2(\omega)$ and $\varphi_2(\omega)$. The spectra were normalized by a reference spectrum acquired from a clean gold surface (100 nm Au-thickness sputtered on silicon). To obtain the far-infrared measurements, we emploied a KRS-5 beamsplitter and a customized Ge:Cu photoconductor, which provided broadband spectral detection down to 320 cm$^{-1}$.

### 2.1.3 Point defects and impurities characterization

For the EPR measurements, , we used a custom build commercial Magnettech MiniScope MS 400 X-band spectrometer coupled to water-cooled electromagnets capable of producing magnetic fields up to 800 mT. The EPR angular dependence at room temperature of a bulk clinochlore sample coupled to a goniometer was performed with *c* axis perpendicular to the rotating axis, in which the *c* axis will be parallel to the applied magnetic field for a specific orientation. The parameters used to acquire all spectra were 9.441(1) GHz with approximately 30 mW of power, 100 kHz of modulation and 0.5 mT of field modulation. The EPR signals are labeled by their effective *g* values defined by the relation $h\nu = g\beta B$, where *h* is Planck's constant, *v* is the microwave frequency, *β* is the Bohr magneton and *B* is the modulus of the magnetic field. The Mössbauer spectrum was measured at 300 K in constant acceleration mode (triangular velocity) using a $^{57}$Co source in Rh matrix. Sample powder was pressed into a sample holder (Ø=0.9 cm) to produce an area of ~0.6 cm$^2$. Due to the lamellar clinochlore structure, the powder presents some expected preferred orientation of the micro flakes when pressed, with their planes tending to be perpendicular to the gamma-ray. Sample thickness was close to the thin absorber approximation, to avoid thickness effects. The spectrum was fitted with Lorentzian doublets using Normos software [47]. The isomer shifts (δ), quadrupole splittings (Δ) and widths (Γ) were allowed to vary to adjust correctly the spectrum. The absorbance and the reflectance spectrum needed to produce the Tauc plot were acquired with a Shimadzu 3600 Plus UV/Vis/NIR spectrometer for a suspended clinochlore flake. The grating change occurs at 720 nm for the spectral range from 300 to 1200 nm during the experiment. XRF hyperspectral mapping was recorded at the Carnauba beamline [48] with synchrotron radiation in panoramic views with a total area of 500 µm x 500 µm with 5 µm pixel size. The X-ray beam spot size is down to 500 nm x 200 nm and the measurements were performed with 9750 eV of excitation energy, with an energy resolution of $\Delta E/E = 10^{-4}$. The images were normalized to the most intense pixel signal using Python scripts and plotted according to the main trace components found in the sample (Fe, Cr, Mn). X-ray absorption near edge structure (XANES) measurements were performed around the Fe K-edge energy in fluorescence mode and room temperature also at Carnauba beamline with nanometric beamsize (down to 500 x 200 nm). The XANES experimental data was normalized and processed using the Athena software from the Demeter suite [49]. For the qualitative data comparison, we used X-ray absorption spectra of standards references, in which Fe can be found in different oxidizing states (metallic Fe – Fe$^0$, FeO – Fe$^{2+}$, Fe$_2$O$_3$ – Fe$^{3+}$ and Fe$_3$O$_4$ – mixed Fe$^{2+}$, Fe$^{3+}$) that were previously recorded at XAFS2 beamline at the UVX synchrotron light source in transmission mode [50].

**2.2 Density Functional Theory (DFT)**

Our ab initio calculations were performed based on the density functional theory (DFT) [51] as implemented in the SIESTA program [52]. The generalized gradient approximation (GGA) within the Perdew-Burke-Ernzerhof (PBE) [53] and vdW-BH [54] parameterizations, which allows a self-consistent treatment of van der Waals interactions, were chosen for the exchange-correlation functionals. We employed norm-conserving Troullier-Martins pseudo-potentials in the factorized Kleinman-Bylander forming a basis set composed of double-zeta pseudo-atomic orbitals of finite range augmented by polarization functions (DZP basis set). Spin polarization was included in all calculations. A real-space grid was used with a mesh cutoff of 350 Ryd and the Brillouin zone was sampled using a k-grid cutoff of 20 Å [55]. For PDOS calculations, the Brillouin zone was sampled using a 4x4x10 k-grid Monkhorst-Pack [56]. All geometries were optimized to obtain a maximum force on any atom less than 10 meV/Å. The optical absorption spectra were calculated using dipole transition matrix elements between Kohn-Sham states with a grid of 20x20x20 k-points for energies up to 10.0 eV. A Gaussian broadening with width of 0.05 eV was used in the absorption spectra. Formation energies were calculated in order to determine the energy cost of substitutional atoms (Cr, Mn and Fe) in the octahedral aluminum sites. The equation used was:

$$E_f = E_{cell} - E_{pristine} + \mu_{Al} - \mu_M$$

where $E_{cell}$ and $E_{pristine}$ are the total energies of the large unit cell that includes the defect and the pristine case, $\mu_{Al}$ is the chemical potential calculated from the total energy of a perfect FCC Al crystal and $\mu_M$ is the chemical potential of the substitutional transition metal. We choose the FCC structure for Mn and BCC for Cr and Fe.

## 3. Results and Discussion

**3.1 - Structural and morphological characterization**

A structural characterization of our natural clinochlore sample extracted from Minas Gerais/Brazil geological environment is a mandatory step prior to the discussions concerning its physical properties. The lamellar structure of clinochlore shown in Fig. 1a consists of a tetrahedral silicon oxide layer, in which an aluminum (Al) atom occupies one of the four silicon (Si) sites, intercalated with two types of octahedral layers [32]. One of the two octahedral layers consists of bivalent magnesium (Mg) ions as central atoms and the other layer is formed by both trivalent Al and bivalent Mg ions as central atoms of the octahedron with hydroxyl (OH) groups at the vertices [10]. The charge balance of the structure is preserved by the presence of some trivalent cation impurities such as chromium (Cr) and iron (Fe) [10] in octahedral sites. Such geometrical disposition provides the excess of positive charge needed to balance the negative charge due to Al substitution at Si sites and unshared oxygen atoms (O) in the formation of the tetrahedral layer. To confirm the crystal structure of our clinochlore mineral, we perform powder XRD measurements combined with a phase-matching throughout refinement using the MAUD package [57] (Fig. 1b). From the analysis, we obtain a complete structural description of the clinochlore sample investigated in this work (Fig. 1c). The most simplified combination of two phyllosilicate minerals that could provide the best fitting to the experimental data that matches peak intensities intensities and positions is retrieved for a volumetric combination of 91(1)% clinochlore and 9(1)% vermiculite, where the amount of vermiculite alteration may be due to a natural weathering or hydrothermal action [10]. The parameters retrieved for the clinochlore phase are $a = 5.3488(6)$ Å, $b = 9.5284(5)$ Å, $c = 14.4241(5)$ Å, $\alpha = 90.93(1)°$, $\beta = 97.72(1)°$ and $\gamma = 88.09(1)°$, leading to a composition and site occupancy compatible with the parameters obtained by Philips and co-workers [58], while the vermiculite parameters were held according to Calle and colleagues [59]. The triclinic (P-1 space group) structure of clinochlore represented in Fig. 1a is generated by the retrieved parameters from the refinement, where the $c$ axis points along the elongated unit cell direction. The lamellar atomic structure of the studied clinochlore sample is revealed to be formed by sheets of tetrahedral sites of Si with 25% of Al substitution stacked along the $c$ axis (with O at the vertices) intercalated by sheets of octahedral sites of pure Mg and sheets of octahedral mixed sites of Mg and Al. In the octahedral Al sites, there is also a 25% of Cr occupation. In the vertices of the octahedral sites, OH groups are present.

To further study the composition of our crystal (Fig. 1c), we perform EDS and WDS analysis (see methods for details). Fig. 1d shows a representative EDS spectrum. An EDS elementary qualitative analysis, using data from ten distinct points of the bulk clinochlore sample, confirms the XRD model through the identification of the constituent elements (O, Mg, Si and Al) with the presence of Cr. EDS measurements also reveal a significant existence of Fe impurities, which are commonly reported to occur in phyllosilicate minerals as substitutional ions at octahedral sites [60]. Quantitative analysis by WDS provide a content of 31.4(3)% of $SiO_2$, 31.0(2)% of $MgO$, 16.8(3)% of $Al_2O_3$, 6.8(4)% of $FeO$ and 0.6(1)% of $Cr_2O_3$, given a total of 86.6(6)%, while the remaining contribution is attributed to the presence of water/hydroxyl ions in the clinochlore mineral. Fe impurities are expected to be located at any octahedral site in the clinochlore structure [60], while Cr ions are expected to occupy only in 25% of the Al-octahedral sites according to the XRD analysis. From the quantitative analysis by WDS, we confirm that the content of Fe impurities is very relevant, while Cr impurities are present in a less expressive amount in our clinochlore sample.

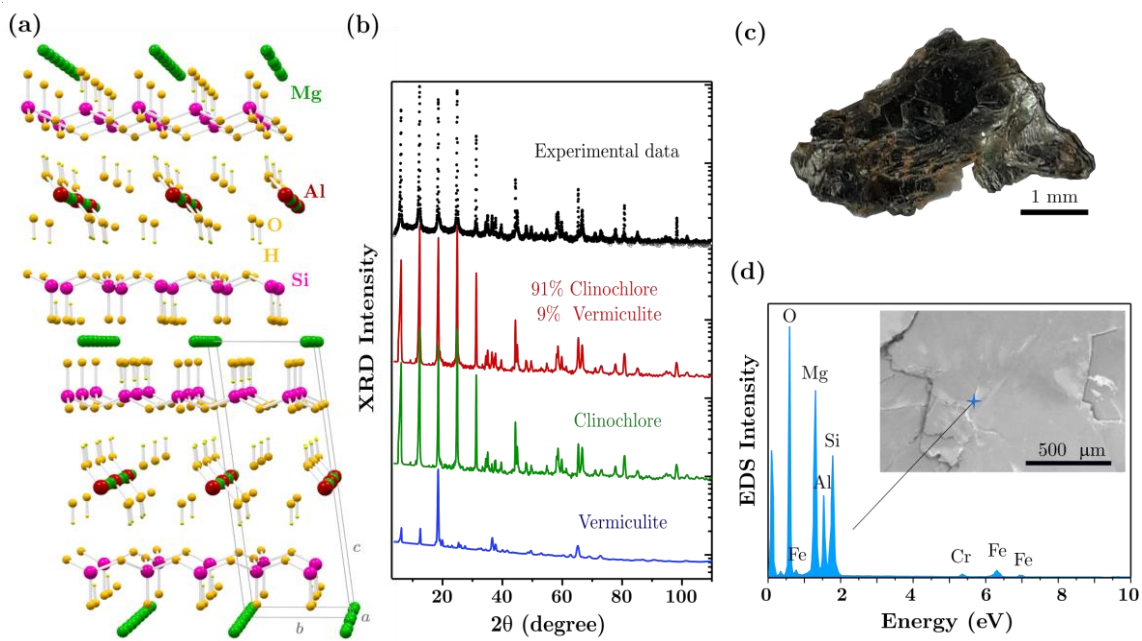

**Figure 1** – The lamellar atomic structure (a) of clinochlore is formed by layers of tetrahedral sites of Si (pink spheres) with 25% of Al substitution stacked along the *c* axis intercalated by layers of octahedral sites of Mg (green spheres) and layers of octahedral sites of Mg and Al (red spheres). In the vertices, O and OH are represented by the large and small yellow spheres, respectively. The XRD measurements and phase matching (b) for the clinochlore sample (c) show that the experimental data (black dots) are well fitted with two phases by refinement using the MAUD package (red line) with 91(1)% contribution of clinochlore (green line) and 9(1)% of vermiculite (blue line). EDS measurements (d) confirm the presence of the constituent elements, and also reveal a content of Cr and Fe impurities.

### 3.2 - DFT calculations and impurity effects on band structure

In order to evaluate the consequences of substitutional transition metals (TMs) on the electronic and optical properties of clinochlore, first-principles DFT calculations (see methods for details) are performed for several possible configurations of the clinochlore structure (Fig. 1a) with substitutional TMs atoms such as Fe, Mn, and Cr in the unit cell. A pristine structure of bulk clinochlore (without substitutional TMs atoms) is geometrically constructed from the experimental data provided by the XRD analysis, which contains 72 atoms. However, with the purpose of ensuring a 25% Al-substitution in the Si-tetrahedral sites (talc-like) and subsequently a 25% TMs-substitution in the Al-octahedral sites (brucite-like), a replication of this structure is built. The final pristine structure in such condition contains 20 atoms of Mg, 32 of H, 72 of O, 12 of Si and 8 of Al, totalizing 144 atoms. Four Al atoms are considered as central atoms of the octahedral sites, while the remaining octahedral sites are occupied by Mg atoms, and the other four Al atoms are considered substitutional atoms at the Si tetrahedral sites. The octahedral substitution by Al atoms ($Al_{Mg}$) acts as a single donor site, compensating the Al tetrahedral substitution ($Al_{Si}$) that acts as a single acceptor site. The resulting electronic structure of the pristine clinochlore is a closed-shell insulator with 0 total spin and a direct band gap of 4.35 eV calculated with the Perdew–Burke-Ernzerhof (PBE) exchange-correlation potential [53] and 4.40 eV calculated with the van der Waals Berland-Hyldgaard (vdW-BH) exchange-correlation potential [54]. It is important to note that bandgap values calculated with DFT are commonly underestimated relative to experimental values [61].

Starting from the pristine structure, we have also considered configurations where a TM atom (Fe, Mn or Cr) replaces an Al atom in the octahedral site of the brucite-like layer. This leads to $Fe^{3+}$, $Mn^{3+}$ and $Cr^{3+}$ sites with a calculated total spin of *5/2, 2,* and *3/2*, respectively. The structural parameters of these configurations calculated with generalized gradient approximation (GGA) PBE and with vdW-BH exchange-correlation functionals are shown in Table 1 along with the experimental parameters obtained by the XRD analysis, showing a good agreement with previously theoretical works [33].

**Table 1** – Clinochlore lattice parameters calculated with different exchange-correlation functionals for the pristine structure and for the structure with TMs substitutional atoms at Al-octahedral sites along the parameters obtained by the XRD analysis in comparison with a previous DFT work in literature.

| Clinochlore Structure | $a$ (Å) | $b$ (Å) | $c$ (Å) | $\alpha$(°) | $\beta$(°) | $\gamma$(°) |
|---|---|---|---|---|---|---|
| **Pristine** | | | | | | |
| GGA-PBE | 5.392 | 9.345 | 14.471 | 90.45 | 97.31 | 89.94 |
| vdW-BH | 5.346 | 9.264 | 14.363 | 90.60 | 97.34 | 89.96 |
| Literature (B3LYP) [33] | 5.306 | 9.188 | 14.336 | 90.18 | 97.06 | 90.00 |
| **Pristine + 1 Fe$_{Al}$** | | | | | | |
| GGA-PBE | 5.404 | 9.365 | 14.478 | 90.41 | 97.35 | 89.94 |
| vdW-BH | 5.357 | 9.282 | 14.369 | 90.64 | 97.38 | 89.97 |
| **Pristine + 1 Mn$_{Al}$** | | | | | | |
| GGA-PBE | 5.408 | 9.362 | 14.469 | 90.53 | 97.19 | 89.78 |
| vdW-BH | 5.359 | 9.277 | 14.360 | 90.81 | 97.21 | 89.81 |
| **Pristine + 1 Cr$_{Al}$** | | | | | | |
| GGA-PBE | 5.401 | 9.361 | 14.473 | 90.44 | 97.31 | 89.94 |
| vdW-BH | 5.345 | 9.275 | 14.369 | 90.68 | 97.38 | 89.96 |
| | | | | | | |
| XRD analysis | 5.349 | 9.528 | 14.424 | 90.93 | 97.72 | 89.09 |

The formation energies of the substitutional TM defects were also calculated to estimate the energetic cost of exchanging an Al atom for a TM in the octahedral site of the brucite-like layer. The values obtained with the GGA-PBE (vdW-BH) approximation are 3.45(3.83) eV for $Fe^{3+}$, 3.15(3.12) eV for $Mn^{3+}$ and 1.80(1.53) eV for the case of $Cr^{3+}$, indicating that the Cr/Al substitution is most favorable. However, it is important to emphasize that the Fe abundance in the natural environment of the mineral formation can favor its incorporation as Fe inclusions and also overcome the higher calculated energy cost for the specific Fe/Al substitution considered. Moreover, Fe ions are frequently present in higher concentration than Mn or Cr in the clinochlore structure as a substitutional impurity at the Al octahedral sites [60], in agreement with our quantitative WDS analysis.

The calculated clinochlore band structures for the TMs substitutions and the densities of electronic states at the $Fe^{3+}$, $Mn^{3+}$, and $Cr^{3+}$ sites are shown in Fig. 2a-c, respectively. The imaginary part of the dielectric function of the $Fe^{3+}$, $Mn^{3+}$, and $Cr^{3+}$ sites are shown in Fig. 2d-f, respectively, which is related with the mineral optical absorption properties and can be taken as a semi-quantitative guide for the interpretation of the experimental results.

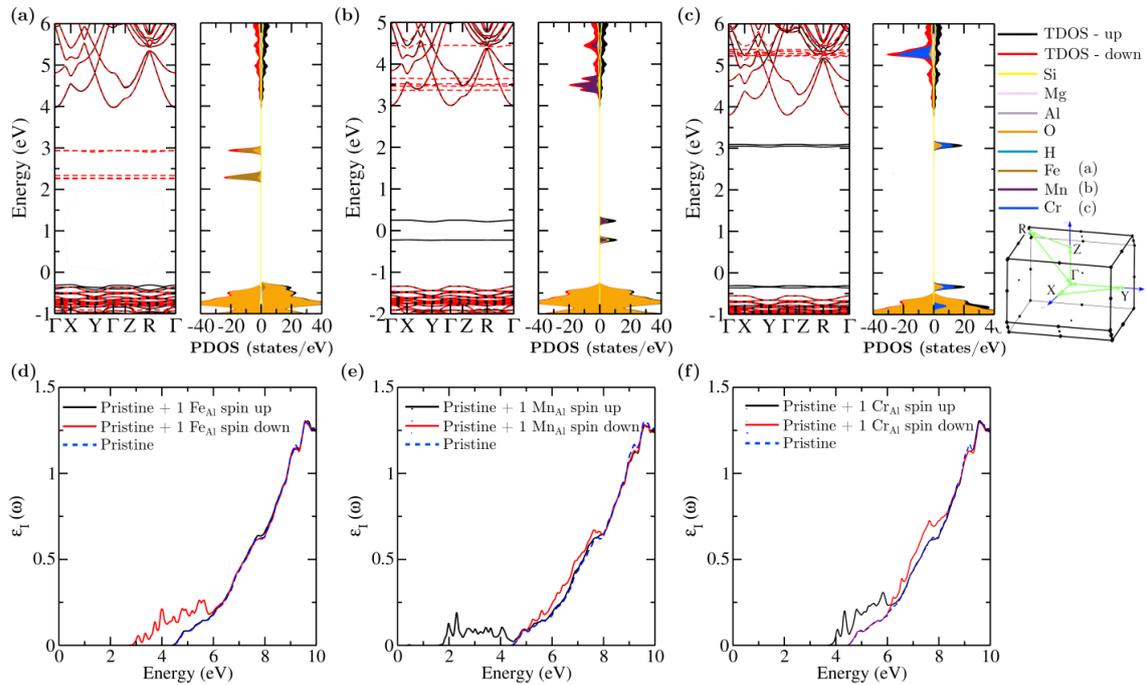

**Figure 2** - Calculated band structures and the densities of electronic states of the (a) $Fe^{3+}$, (b) $Mn^{3+}$ and (c) $Cr^{3+}$ clinochlore sites. Both the total (TDOS) and partial (PDOS) densities of states are shown, the latter projected at atomic sites. All energies are relative to the Fermi level. The calculated imaginary part of the dielectric function of the (d) $Fe^{3+}$, (e) $Mn^{3+}$ and (f) $Cr^{3+}$ sites, which is related with the mineral optical absorption properties.

The electronic structure associated to Fe sites (Fig. 2a), the most abundant TM impurity in the studied sample, shows that $Fe^{3+}$ originates three unoccupied minority-spin impurity states (which appear as flat bands) in the bandgap region. The partial density of states (PDOS) projected at atomic sites indicates that these impurity states are predominantly localized at the Fe atom. From the calculated optical absorption spectrum for $Fe^{3+}$ and the calculated spectrum of the pristine structure (both shown in Fig. 1d), we observe that $Fe^{3+}$ gives rise to optical absorption features in the visible and ultraviolet (UV) range which are non-existent in the pristine structure. These Fe-induced optical absorption features originate from electronic transitions from occupied states near the top of the valence band (VB), which have predominantly O atom character, to the unoccupied impurity states, which have predominantly Fe atom character. Therefore, these features correspond to $Fe^{3+} \rightarrow Fe^{2+}$ charge-transfer optical transitions.

Finally, analyzing the electronic structures associated to the Mn (Fig. 2b) and Cr (Fig. 2c) sites (Fig. 2c) we find that, differently from $Fe^{3+}$, the $Mn^{3+}$ and $Cr^{3+}$ ions originate majority-spin impurity states in the bandgap region. The projected DOS at atomic sites indicate that $Cr^{3+}$ impurity states are more localized at the TM atom than those of $Mn^{3+}$, in accordance with the fact that the Mn ions show a larger oxygen-atom character and this suggests less localized impurity states. The spectra shown in Fig. 2e-f reveals optical absorption features in the visible and UV range for Mn, and in the UV range for Cr ions. Similar to the Fe case described above, these features correspond to $Mn^{3+} \rightarrow Mn^{2+}$ and to $Cr^{3+} \rightarrow Cr^{2+}$ charge-transfer optical transitions. A common feature of all calculated absorption spectra is the existence of structures in the 8-10 eV range, which are related to VB - conduction band (CB) bulk transitions.

## 3.3 - Point defects and impurities characterization

Fe ions can be present in the clinochlore structure with different valence states, i.e., $Fe^{2+}$ and $Fe^{3+}$. $Fe^{2+}$ ions with $(3d^6)$ electronic configuration are non-Kramer ions, being generally unobservable by EPR [62,63]. However, $Fe^{3+}$ $(3d^5)$ ions with an effective electronic spin $S=5/2$ has been extensively investigated for several minerals, where $Fe^{3+}$ ions are commonly present [60,64–66]. Fig. 3a shows the EPR angular dependence obtained at X-band microwave frequency and at room temperature for the bulk monocrystalline clinochlore sample, for rotation of the sample with the $c$ axis perpendicular to the rotation axis. The EPR spectra reveal an anisotropic principal line with an effective g value varying between g~4 and g~2. Some unresolved anisotropic lines at lower fields are also observed. For the angular dependence of EPR spectra with the rotation axis parallel to the $c$ axis (not shown), only a broad unresolved isotropic signal around g~4 is observed. The EPR measurements suggest that the $Fe^{3+}$ ions are situated in the expected octahedral sites, but distorted by a strong rhombic crystal field, commonly reported in the literature and expected for the low-symmetric clinochlore structure [60,65–67]. Previous works also report an isotropic EPR line at g~2 for non-structural Fe ions [65,67], which are not observed in our spectra, indicating that the Fe ions present in our clinochlore sample consist mainly of substitutional ions in distorted octahedral sites. In addition, a broad line at g~2 expected due to the $Cr^{3+}$ contribution [64] is not observed in our EPR spectra, but this signal may have been suppressed by the significant presence of Fe. It is important to note that each possible EPR transition is dependent on the orientation of the applied magnetic field with respect to the symmetry axes of the crystal field [63]. The paramagnetic line broadening of our EPR measurements may be associated with the superposed contribution of paramagnetic centers differing so slightly due to the non-aligned stacking of our clinochlore lamellar structure so that their particular signals cannot be resolved [66]. Another possibility for the occurrence of line broadening in EPR spectra may be due to the exchange between the Fe ions.

Fig. 3b shows the EPR spectra at the specific orientations 180º and 90º of the angular dependence. Despite the non-sensitivity of the EDS and WDS techniques to identify the presence of Mn in the sample due to its lower content (< 0.1%), the presence of Mn in the sample is evident from the EPR spectrum at 90º orientation. The inset in Fig. 3b highlights the observation of six well-resolved lines with the same intensity and spacing that arise from the hyperfine splitting of the paramagnetic $Mn^{2+}$ ions and the $^{55}Mn$ isotope with nuclear spin $I=5/2$ and 100% of natural abundance. The EPR signal attributed to the existence of Mn impurities in the clinochlore structure is present isotropically in all orientations of the angular dependence, but for some orientations, such as 180º, the intensity of the EPR signal associated with Fe impurities dominates the spectrum in the g~2 region and we cannot properly observe the Mn hyperfine lines because EPR intensities are proportional to the content of paramagnetic centers in the material structure [62].

Since EPR measurements are unable to detect the presence of $Fe^{2+}$ ions, Mössbauer absorption measurements are performed (see methods for details). Table 2 lists the hyperfine fitting results for the isomer shift, quadrupole splitting, full width at half maximum, and absorption areas of the two doublets. Fig. 3c shows the corresponding fitted Mössbauer spectrum. The major doublet (red line) is assigned to $Fe^{2+}$ and the smaller doublet (blue line) to $Fe^{3+}$. The $Fe^{2+}$ doublet exhibits intensity asymmetry of its two lines, which is due to the preferred orientation of the micro-flakes relative to the Mössbauer setup. This result indicates that the direction of the electric field gradient must be perpendicular to the plane of the micro-flakes (or parallel to the incident gamma-ray). From the absorption areas of the two doublets, 69.5% for $Fe^{2+}$ and 30.5% for $Fe^{3+}$, we can conclude that the Fe impurities enter the clinochlore structure preferentially in the bivalent state.

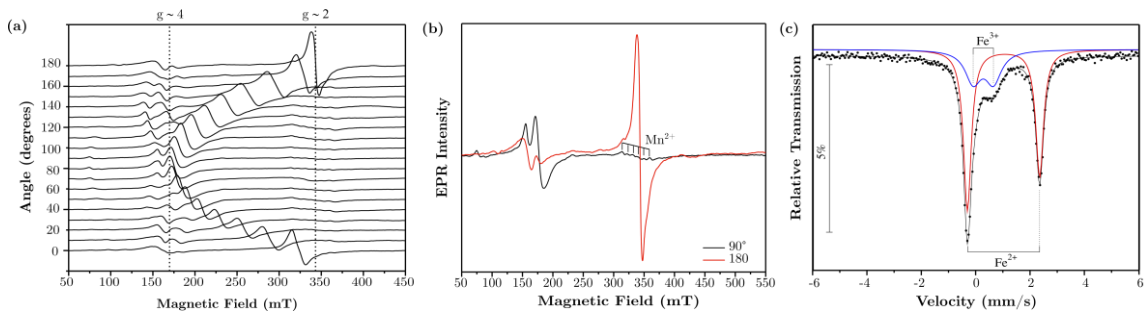

**Figure 3** – The angular dependence of clinochlore EPR spectra at 9.436(1) GHz with the *c* axis perpendicular to the rotation axis (a) indicates a typical dependence of $Fe^{3+}$ ions in an orthorhombic environment. (b) The EPR spectra at the specific orientations 180° (B//c) and 90° from the angular dependence. The presence of Mn in the sample is evident from the spectrum at the specific orientation 90º by the appearance of six well-defined peaks. (c) Mössbauer spectrum of clinochlore fitted with $Fe^{2+}$ (red line) and $Fe^{3+}$ (blue line) doublets contributing with 69.5% and 30.5% respectively.

**Table 2** – Mössbauer hyperfine parameters after fit procedure for the two Fe doublets, with respective values of isomer shift ($\delta$), quadrupole splitting ($\Delta$), width ($\Gamma$) and area. Errors of the hyperfine parameters ($\delta$, $\Delta$, $\Gamma$) are estimated as ±0.01 mm/s and ±0.02 mm/s for $Fe^{2+}$ and $Fe^{3+}$, respectively.

| Doublet | $\delta$ (mm/s) | $\Delta$ (mm/s) | $\Gamma$ (mm/s) | Area (±0.1%) |
|---|---|---|---|---|
| $Fe^{2+}$ | 1.13 | 2.65 | 0.35 | 69.5 |
| $Fe^{3+}$ | 0.39 | 0.75 | 0.70 | 30.5 |

Optical experiments in X-ray/UV/Vis/NIR spectral region are then performed to investigate the optical assignments of TMs substitutional ions, determine the experimental bandgap of the material, and characterize the high-energy (> 4 eV) structures in the calculated absorption spectra given by the first-principles DFT analysis. Starting from the UV/Vis/NIR absorbance spectrum of a suspended clinochlore flake (Fig. 4a), we observe different bands (peaks) associated with Fe impurities in both $Fe^{2+}$ and $Fe^{3+}$ valence states. The absorption bands observed below 600 nm are related to orbital electronic transitions of $Fe^{3+}$ ions, while the two broad bands at ~900 nm and ~1100 nm are due to the presence of $Fe^{2+}$ ions [60]. The absorption band observed at ~700 nm in the middle of the explored spectral range was previously attributed to the charge transfer between iron ions [60]. Absorption bands associated with $Cr^{3+}$ ions, which are expected to be observed around 430 nm and 610 nm [60], may have been suppressed by Fe absorption bands. This is in agreement with our WDS quantitative analysis that reveals that the Fe content is much more relevant than the Cr amount. In order to experimentally determine the optical bandgap of the studied clinochlore sample, a Tauc plot (inset in Fig. 4a) is elaborated from a reflectance measurement performed in the same setup of the absorbance measurement described above. The bandgap is obtained by extrapolation of the linear behavior (red dashed line) of the Kubelka-Munk function of our data [68,69]. Considering a direct bandgap for the clinochlore, we obtain an experimental band gap of 3.6(1) eV. The bandgap value suggested by the DFT calculations for the pristine clinochlore structure is about 4.4 eV, while the calculations considering different TM ion substitutions show a reduction of up to half of the effective optical gap energy due to the insertion of defect states in the bandgap region. This is in good agreement with the experimental bandgap 3.6(1) eV retrieved. We must mention that in a realistic scenario, the mineral has various TMs substitutions that mutually contribute to the reduction of the effective bandgap energy compared to the pristine structure, besides the well-known underestimation of the bandgap energy obtained by DFT calculations.

We now focus our attention to the DFT calculated features that appear in the range of 8-10 eV (Fig. 2d-f). For this purpose, we perform UR measurements in our bulk-crystal (see methods for details). The representative UR spectrum shown in Fig. 4b confirms the existence of optical transitions between the VB and the CB calculated by DFT, as can be seen by the similarity between the position of the peaks demarcated in the UR spectrum and those of the perceived structures in the spectra in Fig. 2d-f.

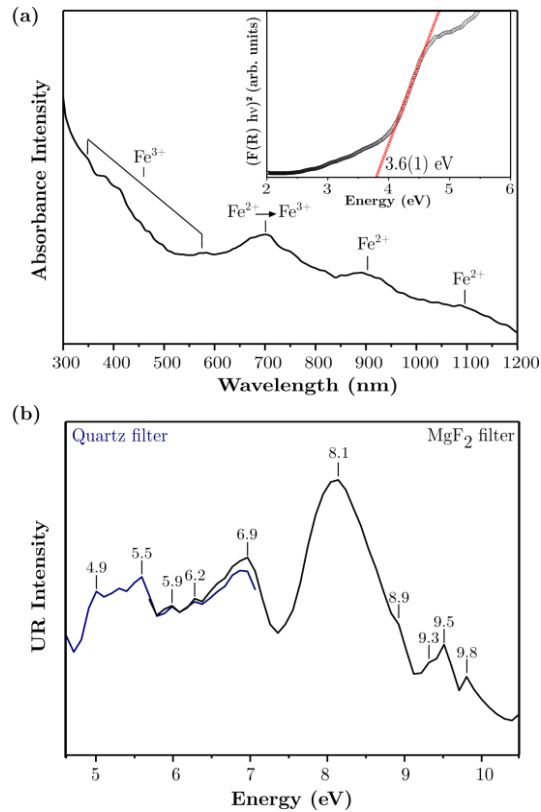

**Figure 4** – The UV/Vis/NIR absorbance spectrum (a) of a suspended clinochlore flake confirms the presence of Fe impurities in different valence states. The inset Tauc plot in (a) shows an experimental estimation of 3.6(1) eV for the clinochlore optical bandgap. (b) Clinochlore UR measurements revealing several peaks related to transitions between the valence and conduction bands.

It is important to mention that due to the high Fe content in the clinochlore sample studied, it is possible to analyze this impurity by various techniques, such as EPR, Mössbauer and UV/Vis/NIR absorption. However, little experimental information could be obtained for Mn and Cr impurities by the complementary techniques already discussed. The quantification obtained by WDS was capable to point out traces of Cr in the sample, but the allowed absorption bands in the visible region associated with Cr ions could not be properly identified, most likely due to the predominant Fe bands in the spectrum in Fig. 4a. EPR measurements clearly show the Mn spectral signatures, confirming the presence of this impurity in the sample, but no trace of Mn was quantified by WDS analysis, indicating the amount of Mn in the clinochlore sample is not expressive. In an attempt to clarify this discussion, XRF measurements are performed on a 500 μm x 500 μm panoramic clinochlore area (see methods for details), resulting in the impurity maps shown in Fig. 5a-c with normalized intensity for each investigated metal (Fe, Mn, and Cr). We can observe from these XRF maps that the Mn and Cr impurities are homogeneously distributed over the investigated area (Fig. 5a-b, respectively), suggesting that they are preferentially incorporated into the clinochlore structure. On the other hand, the XRF map for Fe impurities (Fig. 5c) reveals an inhomogeneous distribution over the sample area, forming Fe inclusions. From our previous and extensive analysis of Fe impurities, it is known that the Fe ions are predominantly bivalent by Mössbauer analysis, while EPR measurements determine that $Fe^{3+}$ ions are present in the sample is distorted octahedral sites and the absence of an isotropic line at g~2 excludes the possibility that trivalent Fe ions are present in the sample at non-structural sites in significant amounts. For these reasons, our results suggest that Fe inclusions in the clinochlore structure should preferably be constituted by $Fe^{2+}$ ions, while the smaller amount of $Fe^{3+}$ may be homogeneously incorporated into the clinochlore structure, as well as the Mn and Cr ions.

To explore the possibility that Fe inclusions are formed mainly by bivalent ions, we performed XANES measurements in a nanometric region (see methods for details) enclosed by the black circle in Fig. 5c. The respective XANES spectrum recorded around the Fe K-edge energy is plotted in Fig. 5d in qualitative comparison with standard reference data of iron oxides presenting Fe in different valence states. Comparatively, the experimental data is well described by the XANES spectrum of the FeO standard over the others, corroborating the idea suggested by our results that the constitution of the inclusions is dominated by bivalent Fe ions. This analysis is also compatible with the fact that Fe has the highest energy cost, calculated by our DFT, to replace an Al atom in the octahedral sites, favoring its incorporation into the material structure mostly as segregated inclusions. It is important to stress that previous works demonstrated that heat treatment and acid leaching can remove some impurities from natural phyllosilicates [70,71], thus, this process could potentially be applied to our material to purify and further increase the bandgap energy (at least up to 4.4 eV, as suggested by our DFT calculation), restoring some of the pristine characteristics of the crystal (these suggested processes are out-of-the scope of the current work). This is desired for applications using clinochlore as a suitable and alternative 2D insulating material, highly needed for the development of nanoelectronic devices [25].

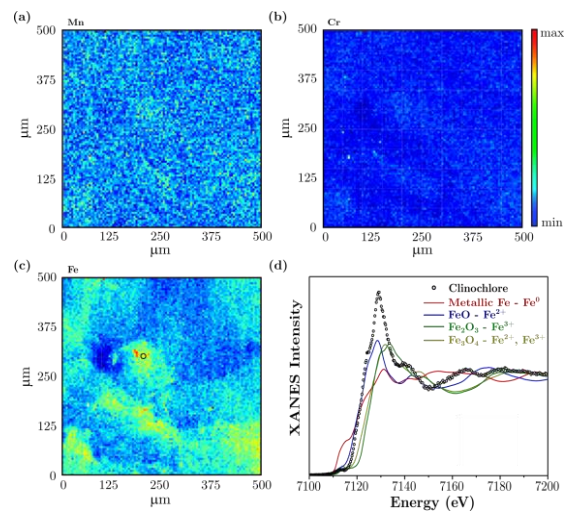

**Figure 5** – Fluorescence normalized intensity hyperspectral maps acquired under 9750 eV X-ray synchrotron excitation for the transition metal impurities (a) Mn, (b) Cr and (c) Fe present in the clinochlore sample. (d) Nanometric XANES spectrum recorded at the black circle delimited region in (c) around Fe K-edge energy in comparison with standard iron oxides presenting iron in different valency states (solid lines).

**3.4 - Vibrational characterization**

In order to investigate possible modifications in the molecular-vibrational identity of clinochlore due to the presence of impurities or structural defects, the sample is investigated by Raman and FTIR spectroscopy. The non-destructive analytical power microscale enables these techniques to be extremely useful to study LMs [72–74]. The Raman spectrum of a bulk clinochlore sample shown in Fig. 6a is in close agreement with the spectrum calculated by Ulian and co-workers [75], as well as to previous studies in natural crystals [60,76–78]. The spectrum features 10 strong and well-defined modes in the range of 200 to 1100 $cm^{-1}$, while OH modes lie between 3500 and 3750 $cm^{-1}$. The OH-stretching bands in this region are attributed to hydrogen-bonded interlayer OH [76,77]. The peaks around 1089 and 1059 $cm^{-1}$ are attributed to antisymmetric Si-O-Si stretching modes [77,78], while the symmetric Si-O-Si stretching appears as strong bands at ~687 and 553 $cm^{-1}$ [60,76–78]. The peaks at 459 and 440 $cm^{-1}$ are probably assigned due to librational OH and bending Si-O-Si modes [78]. The strong band at 353 $cm^{-1}$ has been assigned to the bending vibration of Si-O [60] and its intensity decreases with the Si substitution by Al [77]. The peak at 283 $cm^{-1}$ is probably related to movements of the tetrahedral sheet, whereas the sharp peak at 203 $cm^{-1}$ was ascribed to vibrations of the octahedrons [75,78]. Finally, the peak at 125 $cm^{-1}$ is regarded as translational Si-O-Si modes [60,75–78]. It is worth mentioning that the two other weaker peaks observed at around 106 $cm^{-1}$ and 158 $cm^{-1}$ are related to bending Si-O and to a combination of the out-of-plane vibration of the Mg ions with librations of the OH groups in the brucite-type interlayer, respectively.

The clinochlore FTIR absorbance spectrum is shown in Fig. 6b. The strong bands and shoulders registered in the 900 - 1100 $cm^{-1}$ spectral region are assigned to stretching vibrations of the tetrahedral $SiO_4$. The spectral region between 760 and 850 $cm^{-1}$ could be associated with tetrahedral Al–O stretching, whereas the band at 677 $cm^{-1}$ could be attributed to either symmetric Si–O–Si stretching mode or metal-hydroxyl libration. The band observed at 1627 $cm^{-1}$ of the –OH bending indicates strong coordination of the metal ion with the surrounding –OH groups. The absorption bands within 3000 and 3750 $cm^{-1}$ are specifically related to OH group and the three main bands observed have been assigned to stretching vibrations of OH functional group [60]. In particular, the band at 3438 $cm^{-1}$ is attributed to the stretching of OH bound to $Fe^{3+}$ ions, confirming the presence of Fe impurities located at octahedral sites already suggested by the previous techniques. In Fe-rich chlorites, these absorption bands usually occur at lower frequencies than 4440 and 4270 $cm^{-1}$, while in Mg-rich chlorites the absorptions usually occur at positions higher than these two frequencies [79]. These two bands are present in the FTIR absorbance spectrum of our clinochlore sample at positions slightly blue-shifted, as expected since our WDS quantitative analysis provide a FeO content of 6.8(4)% against a higher MgO content of 31.0(2)% for our clinochlore sample.

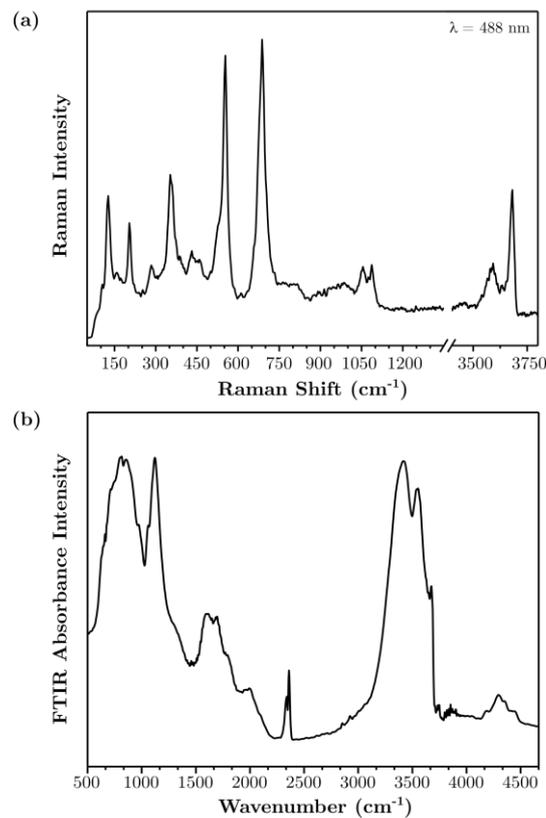

**Figure 6** – Raman spectrum excited at 488 nm (a) and FTIR spectrum (b) of the bulk clinochlore sample.

After the comprehensive molecular-vibrational characterization of the bulk material, one must investigate nanometer-thin 2D flakes produced by mechanical exfoliation. For that, standard scotch tape exfoliation is carried out to obtain clinochlore flakes since the material has weak van der Waals force between layers. A representative optical microscopy image of several exfoliated flakes onto SiO$_2$/Si substrate is shown in Fig. 7a. The apparent color of the flake transferred over 300 nm SiO$_2$/Si under the optical microscope depends on its thickness. Thus, it is possible to identify or estimate the number of layers by their color, which represents thickness variations from a few nanometers up to hundreds of nanometers. The lateral flake size depends directly on the exfoliation process (scotch tape repetition) and it may vary from sample to sample. After material exfoliation, we performed morphological surface analysis by AFM (see methods for details) in a representative clinochlore flake deposited atop the 300 nm SiO$_2$/Si substrate. Similar to other LMs materials, such as as hexagonal boron nitride (hBN) [5], clinochlore shows atomically flat surface over large areas. Fig. 7b shows an AFM topography image measured on a clinochlore flake with approximately 58 nm of thickness and average roughness measured as R$_{rms}$ = 0.26 nm. These results indicate that our samples have a suitable surface topography, making them promising candidates as a substrate or encapsulating media of other LMs in 2D van der Waals heterostructures applications [22,23,80].

Fig. 7c shows another optical microscopy image of an isolated clinochlore flake deposited atop the SiO$_2$/Si substrate. The color contrast in the image indicates a flake thickness varying from 6 nm (green region) up to 200 nm (yellow region). Additionally, a high-resolution AFM topography image is acquired from the highlighted area in the optical image, from which the corresponding profile along stair-like clinochlore layers is measured. The horizontal green line in the high-resolution AFM topography image indicates the extension where the profile of the flake is extracted. The line profile indicates a step height of 6 nm between the substrate and the first adjacent terrace, corresponding to approximately 4 layers of clinochlore, while the next step has a height of 3 nm, corresponding to approximately 2 layers. The results clearly show the existence of few nanometer-thick layers within the clinochlore flake, suggesting that isolated 1L-clinochlore could be eventually produced and identified by AFM. Since the average length of exfoliated clinochlore flakes is approximately 15-30 microns (see scale bar in Fig. 7a and Fig. 7c), the length-to-thickness ratio is ~ 10,000, which formally makes clinochlore a genuine 2D materials at the nanoscale.

An important figure of merit that assures that 2D clinochlore samples are successfully produced is the lack of structural damage and/or process-induced defects that may affect their properties if compared to the bulk material. That is, the spectral response and vibration signature should remain intact in exfoliated flakes. Classical FTIR spectroscopy is unsuitable to access the nanoscale vibrational properties of this LM due to the diffraction limit [81]. Hence, ultramicroscopy modalities are demanded such task. In that sense, SINS enables sub-wavelength spatial resolutions for FTIR by overcoming the diffraction limit and providing the high sensitivity needed to measure extremely small volumes. The SINS technique exploits the high-brightness and ultra-wide coverage of broadband light sources (typical frequency window from $330 - 4000$ cm$^{-1}$ from a synchrotron IR) coupled to FTIR spectrometer combined with an AFM (see methods for details). Therefore, the technique can access the morphology and dielectric response of LMs simultaneously. From the high-resolution AFM topography image in Fig. 7d (left panel), we achieved the thickness of the exfoliated flake in this case ranging between 11 and 20 nm. Simultaneously, the broadband near-field image (right panel) is acquired with sharp optical reflectivity contrast for different amounts of stacked layers, allowing the precise location of candidate positions for SINS point spectra (such as the red and black dots). Fig. 7e shows SINS point-spectra of the clinochlore flake with ~11 and 20 nm of thickness acquired at the red and black dot position in Fig. 7d (right panel), respectively. The measured SINS spectra unveil a strong IR activity in the frequency range from 400-1150 cm$^{-1}$ that matches their micro-FTIR counterparts. Following the results obtained by micro-FTIR, the Si–O in-plane stretching vibration bands have been observed at 1023 and 987 cm$^{-1}$, while the Si–O in-plane bending vibration has been observed at 445 cm$^{-1}$. Al–O stretching is associated with the weak band at 811 cm$^{-1}$ and the two bands observed at 680 and 470 cm$^{-1}$ are assigned to the metal-hydroxyl bond libration and translation, respectively. In the 400–550 cm$^{-1}$ region, the IR spectra displayed three bands near 521, 492, and 420 cm$^{-1}$. These bands are assigned to (Fe, Mg)-O-Si or Al-O-Si bending [77]. Vibrational mode assignments are made via comparison with standard IR spectra, already discussed for the FTIR absorbance spectrum in Fig. 6b.

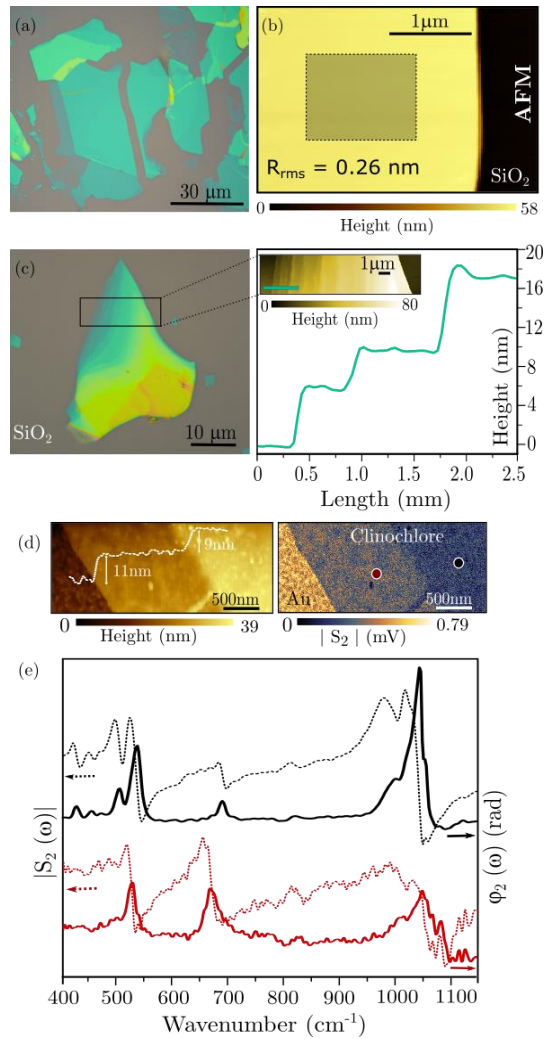

**Figure 7** – (a) Optical microscopy image of several clinochlore flakes exfoliated on a 300 nm SiO$_2$/Si substrate. The different colors of the flakes correspond to different thicknesses. AFM topography image (b) of a clinochlore flake with ~58 nm of thickness and average roughness (R$_{rms}$) acquired in the area delimited by the shaded square mark. From an isolated clinochlore flake (left panel (c)), an AFM topography image of the highlighted region is acquired (insert in right side (c) graph) and the corresponding profile along the green line in the AFM image is taken (right side (c) graph). (d) AFM topography image of a clinochlore FL-flake at a gold substrate with respective AFM profile line (left panel) and broadband near field image of the same region revealing scattering intensity modulation according to the number of layers (right panel). (e) SINS point-spectra: amplitude S$_2(\omega)$ (dashed line) and phase $\varphi_2(\omega)$ (continuous line) of clinochlore flake with 11 nm (red) and 20 nm (black) of thickness acquired at the red and black dot position in (d), respectively. The overall scale focuses on the currently available near-field frequency window (400 - 1150 cm$^{-1}$) of interest.

It is important to compare the SINS point spectrum for the nanoscale flake in Fig. 7d with the FTIR spectra of the bulk material in Fig. 6b. Even though the relative intensities of the main peaks in the 500-1150 cm$^{-1}$ range are different, their position remains the same, indicating the same structural nature. This is a strong evidence that the mechanical exfoliation process does not modify the mineral electronic band structure, therefore, allowing further and more complex studies in 1L or FL-clinochlore samples. Finally, we must mention that all features discussed in our work (large bandgap material, ultrathin layers, atomically flat surface, and air stability) would make clinochlore crystals promising candidates for use in ultrathin and flexible low-cost LM-based optoelectronic devices.

## 4. Conclusion

Clinochlore is one of the most abundant phyllosilicate minerals in nature and in this work, we demonstrated the lamellar character that allows it to be mechanically exfoliated down to a few layers. In particular, we cover up the lack of studies focusing on the exfoliation of 2D structures from clinochlore bulk form with an exploration of its optical properties and the influence of impurities on its macroscopic properties. We performed a robust investigation showing the structural and spectral characteristics of this material. As it is a naturally abundant mineral, the presence of impurities is common in geological samples. We then identified, quantified, and analyzed the transition metal impurities (iron, manganese, and chromium) present in the studied sample through a systematic investigation performed by several complementary experimental techniques, accompanied by a complete density functional theory study. We demonstrated that the incorporation of manganese and chromium impurities takes place homogeneously in the material, while iron atoms can segregate as inclusions. Moreover, our analysis showed that these impurities significantly modify the material properties. Additional optical transitions were observed as well as a reduction in the optical bandgap. Nevertheless, the existence of these impurities could boost its use in defect engineering for localized luminescence experiments of single-photon emission/absorption phenomena. The results discussed in our work (large bandgap material, ultrathin layers, atomically flat surface and air stability) would make clinochlore crystals promising as a stable and optically active component for ultrathin and flexible low-cost LM-based optoelectronic devices.

## Acknowledgments

The authors acknowledge financial support from CNPq, Brazilian Institute of Science and Technology (INCT) in Carbon Nanomaterials, CAPES, FAPEMIG and CNPq. The authors would like to acknowledge the Center of Microscopy, LabCri and LCPNano at the Universidade Federal de Minas Gerais (UFMG) for providing the equipment and technical support for the experiments. The authors also like to acknowledge the Brazilian Synchrotron Light Laboratory (LNLS) and Advanced Light Source (ALS) for the facilities in experiments involving synchrotron radiation and its associated installations. Also, Hans Bechtel for the experimental assistance. I.D.B. acknowledges the support from CNPq through the Research Grant 311327/2020-6 and the prize L'ORÉAL-UNESCO-ABC for Women in Science Prize - Brazil/2021). R.O.F. acknowledges the support from CNPq through the Research Grant 311564/2018-6 and FAPESP Young Investigator Grant 2019/14017-9. A.R.C. acknowledges the financial support through the Fundo Mackenzie de Pesquisa e Inovação (MackPesquisa No. 221017) and the CNPq (No. 309920/2021-3). M.J.S.M. thanks for the support from Universidade Federal de Ouro Preto (UFOP). L.A.G.G. thanks for the support from GCUB-OEA. We also acknowledge computational support from CESUP-UFRGS. All authors thank professor Marco A. Fonseca from UFOP for supplying the clinochlore samples.